\documentclass{article}

\usepackage[preprint]{neurips_2026}

\usepackage[utf8]{inputenc} 
\usepackage[T1]{fontenc}    
\usepackage{hyperref}       
\usepackage{url}            
\usepackage{booktabs}       
\usepackage{amsfonts}       
\usepackage{nicefrac}       
\usepackage{microtype}      
\usepackage{xcolor}         
\usepackage{amsmath}
\usepackage{amsthm}
\usepackage{amssymb}
\usepackage{booktabs}
\usepackage{algorithm}
\usepackage{algorithmic}
\usepackage{soul}
\usepackage{url}
\usepackage{natbib}
\usepackage{xcolor}
\usepackage[most]{tcolorbox}
\usepackage{tcolorbox}
 
\newtcolorbox{takeaway}{
  colback=gray!15,
  colframe=black,
  boxrule=0.6pt,
  arc=2pt,
  left=6pt,
  right=6pt,
  top=4pt,
  bottom=4pt,
  fontupper=\small
}
\usepackage{microtype}
\usepackage{subcaption}

\title{Sparse Autoencoders as a Steering Basis for Phase Synchronization in Graph-Based CFD Surrogates}

\author{%
  Yeping Hu\thanks{Corresponding Author} \quad  Ruben Glatt \\
  Computational Engineering Division\\
  Lawrence Livermore National Laboratory\\
  Livermore, CA 94550 \\
  \texttt{\{hu25, glatt1\}@llnl.gov} \\
  \And
  Shusen Liu\\
  Center for Applied Scientific Computing\\
  Lawrence Livermore National Laboratory\\
  Livermore, CA 94550 \\
  \texttt{liu42@llnl.gov} \\
}

\begin{document}

\maketitle

\begin{abstract}
Graph-based surrogate models provide fast alternatives to high-fidelity CFD solvers, but their opaque latent spaces and limited controllability restrict use in safety-critical settings. A key failure mode in oscillatory flows is phase drift, where predictions remain qualitatively correct but gradually lose temporal alignment with observations, limiting use in digital twins and closed-loop control. Correcting this through retraining is expensive and impractical during deployment. We ask whether phase drift can instead be corrected post hoc by manipulating the latent space of a frozen surrogate. We propose a phase-steering framework for pretrained graph-based CFD models that combines the right representation with the right intervention mechanism. To obtain disentangled representation for effective steering, we use sparse autoencoders (SAEs) on frozen MeshGraphNet embeddings. To steer dynamics, we move beyond static per-feature interventions such as scaling or clamping, and introduce a temporally coherent, phase-aware method. Specifically, we identify oscillatory feature pairs with Hilbert analysis, project spatial fields into low-rank temporal coefficients via SVD, and apply smooth time-varying rotations to advance or delay periodic modes while preserving amplitude-phase structure. Using a representation-agnostic setup, we compare SAE-based steering with PCA and raw embedding spaces under the same intervention pipeline. Results show that sparse, disentangled representations outperform dense or entangled ones, while static interventions fail in this dynamical setting. Overall, this work shows that latent-space steering can be extended from semantic domains to time-dependent physical systems when interventions respect the underlying dynamics, and that the same sparse features used for interpretability can also serve as physically meaningful control axes.
\end{abstract}
\section{Introduction}

High-fidelity computational fluid dynamics (CFD) remains the standard approach for analyzing complex unsteady flows, but its computational cost limits its use in settings that require rapid forecasting, repeated queries, or continual alignment with incoming observations \citep{najm2009uncertainty}. Graph-based surrogate models \citep{pfaff2020learning,hu2023graph, leim4gn} offer an attractive alternative by learning flow evolution directly on the simulation mesh, often at much lower cost than full solvers. However, the node-level embeddings produced by these models are high-dimensional and not directly interpretable, making it difficult to diagnose or correct prediction errors once a rollout begins to deviate. This lack of interpretability and controllability hinders their deployment in safety-critical or regulation-bounded settings \citep{walke2023artificial}, particularly when real-time synchronization with observations is required.

A practically important failure mode in oscillatory flows is \emph{phase drift} \citep{brunton2015closed}. A surrogate rollout may continue to produce qualitatively plausible coherent structures, such as vortex streets or wake patterns, while gradually falling out of synchrony with observations as small errors in phase or frequency accumulate over time \citep{lusch2018deep}. For example, in monitoring flow around a turbine blade, a surrogate may correctly predict the vortex street pattern but progressively lag behind real-time sensor measurements by tens of time steps, rendering its predictions unusable for closed-loop control without expensive model retraining. In this regime, the surrogate has not necessarily learned the wrong dynamics; rather, it generates the right structures at the wrong times. This phase-drift problem is particularly critical in applications such as digital twins for fluid-structure interaction monitoring, real-time flow control systems, and design optimization workflows where temporal alignment between predictions and sensor measurements is essential for downstream decision-making \citep{brunton2015closed}. For such errors, retraining or fine-tuning is a heavy remedy: it changes model weights, requires additional optimization and validation, and is poorly matched to settings where corrections must be applied repeatedly during deployment. This raises a natural question: \textit{can phase drift be corrected post-hoc by manipulating the internal activations of a frozen surrogate during inference, without retraining?} Recent successes in latent-space steering for language and vision models \citep{turner2023steering, zou2023representation} suggest that internal representations encode sufficient structure to enable targeted behavioral adjustments, but whether such techniques transfer to the continuous, time-dependent dynamics of physical systems remains an open question.

If such {latent-space steering} \citep{zou2023representation, turner2023steering, kulkarni2025interpretable} is feasible for CFD surrogates, addressing phase drift requires answering two coupled questions: \textit{(i)~in which representation should the steering be performed?} and \textit{(ii)~what intervention mechanism can correct the phase error without damaging the underlying dynamics?} The choice of representation determines whether oscillatory phenomena can be isolated from other flow physics, enabling targeted edits that remain localized. The choice of mechanism determines whether the edit respects the coupled amplitude--phase structure inherent in time-dependent flows. Answering either question in isolation is insufficient, because even a well-chosen representation cannot compensate for a structurally inappropriate edit, and vice versa.

Regarding the representation question, sparse autoencoders (SAEs)~\citep{Cunningham2023,Gao2024,Marks2024,Mudide2024,Muhamed2024} provide a natural candidate. By training wide, overcomplete, and sparsity-regularized autoencoders on hidden activations, SAEs can discover monosemantic features that correspond to human-understandable concepts~\citep{higgins2017beta,chen2018isolating,locatello2019challenging}. These features form a dictionary of disentangled latent directions, each representing a distinct mechanism within the underlying model. In oscillatory flows, this disentanglement is particularly valuable: if vortex-shedding dynamics can be isolated into a small subset of features with minimal coupling to boundary-layer or pressure-gradient physics, then phase correction can be applied without inadvertently perturbing unrelated aspects of the flow. The role of SAEs in our approach is therefore not merely interpretability but {intervention suitability}: if phase correction requires relatively isolated oscillatory coordinates, then a sparse, disentangled feature basis is a principled space in which to operate.

For the intervention question, existing latent-steering methods from language and vision \citep{subramani2022extracting, li2023inference, turner2023steering, rimsky2024steering, kulkarni2025interpretable, yan2025visual} have shown that internal representations can be manipulated after training to redirect model behavior, typically through static per-feature interventions such as scaling, additive shifts, or clamping \citep{o2025steering}. Phase correction in unsteady CFD, however, is fundamentally different from editing a relatively static semantic attribute. Oscillatory flow features encode spatiotemporal dynamics in which amplitude and phase are temporally coupled: a feature's activation at time~$t$ depends not only on its spatial pattern but also on where it sits within its periodic cycle. Static per-feature interventions such as scaling or additive shifts manipulate amplitude and phase independently, disrupting the coherent temporal organization required to advance or retard a periodic mode. This suggests that effective steering in CFD must be both \textit{phase-aware} (operating on near-quadrature feature pairs that represent sine--cosine decompositions of oscillations) and \textit{temporally coherent} (applying smooth time-varying corrections rather than fixed scalar perturbations).

\begin{figure}[t]
    \centering
    \includegraphics[width=1\linewidth]{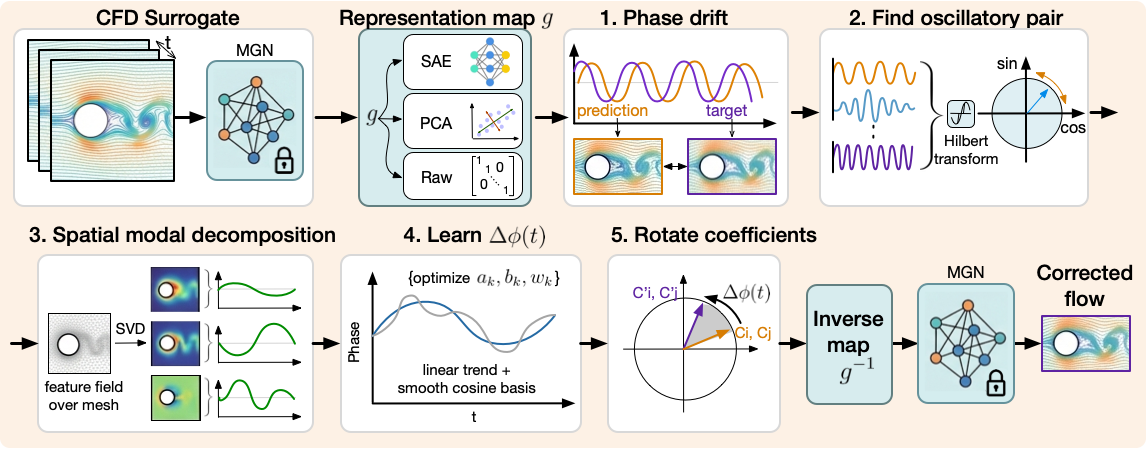}
    \caption{Overview of the phase-steering framework. A pretrained, frozen MeshGraphNet (MGN) produces node embeddings that are passed through a representation map $g$, which can be a sparse autoencoder (SAE), PCA projection, the identity (raw embedding), or any other suitable basis. Given a detected phase mismatch between surrogate predictions and target observations (step~1), oscillatory feature pairs exhibiting near-quadrature relationships are identified via Hilbert transform analysis (step~2). Each selected feature field undergoes spatial mode decomposition via SVD to obtain low-dimensional coefficient trajectories (step~3). A smooth, time-varying phase offset $\Delta\phi(t)$, parameterized by a linear trend and low-frequency cosine basis, is optimized over the steering horizon (step~4). The learned offset is applied by rotating the coefficient pairs, advancing or retarding the oscillation phase without altering its amplitude or spatial structure (step~5). The steered representation is decoded through the inverse map $g^{-1}$ and rolled out through the frozen MGN to produce phase-corrected flow predictions. Only the low-dimensional steering parameters $\{a_k, b_k, \mathbf{w}_k\}$ are optimized. The same pipeline is applied identically in all three representation spaces; only the choice of $g$ differs.}
    \label{fig:framework}
\end{figure}

We address both questions jointly with a unified, representation-agnostic steering pipeline. Given the frozen node embeddings of a pretrained MeshGraphNet~(MGN)~\citep{pfaff2020learning}, we train an SAE and identify latent feature pairs that exhibit matched dominant frequencies and near-quadrature phase relationships---precisely the structure needed for sine--cosine decomposition of periodic modes. For each pair, we compute a low-rank spatial mode decomposition to compress high-dimensional node fields into tractable coefficient trajectories, then optimize a smooth, time-varying phase offset by rotating the pair coefficients over a short prediction horizon. The modified representation is mapped back through the frozen surrogate, yielding phase-corrected predictions without updating any model weights. Because the pipeline is representation-agnostic, it enables a controlled comparison: we apply the same rotation-based steering mechanism in SAE space, PCA space, and the raw MGN embedding, thereby isolating the effect of representation quality on steering performance. Results show that SAE substantially outperforms both alternatives, and that standard static interventions fail entirely in this dynamical setting. Figure~\ref{fig:framework} provides an overview of the complete pipeline, from phase-mismatch detection through oscillatory-pair identification, spatial mode decomposition, phase offset optimization, and rollout through the frozen surrogate. The details are discussed in Section \ref{sec:method}.

Our contributions are as follows:
\begin{itemize}
    \item We formulate phase-drift correction in oscillatory CFD surrogates as a \emph{post-hoc steering problem on a frozen graph-based surrogate}, and introduce a phase-aware, temporally coherent intervention pipeline based on rotations in oscillatory latent subspaces identified via Hilbert analysis.
    \item We provide a \emph{representation-agnostic framework and controlled comparison study} by applying the same rotation-based steering mechanism in SAE space, PCA space, and the raw MGN embedding, demonstrating that SAE achieves $+26.1\%$ fractional MSE improvement versus $+16.0\%$ (PCA) and $+4.1\%$ (raw) under identical conditions.
    \item We show that \emph{standard static latent interventions} (scaling, additive offsets, clamping) do not transfer to this dynamical setting, with performance ranging from zero effect to catastrophic degradation, and demonstrate that effective surrogate steering requires both a sparse, disentangled representation and a structure-preserving intervention design informed by flow physics.
\end{itemize}

\section{Related Work} 
This work touches on several related topics including physics-informed scientific machine learning for surrogate modeling, deep neural network interpretability, and model steering/intervention through latent space.

\subsection{Graph-based Surrogates for Physics Simulation}
Graph-based surrogate models have emerged as powerful alternatives to traditional CFD solvers. \citet{pfaff2020learning} introduced MeshGraphNets (MGN), which achieve state-of-the-art accuracy on unstructured meshes through message-passing neural networks. 
Subsequent work has extended these models to handle multiscale phenomena \citep{fortunato2022multiscale}, multiple physics \citep{sanchez2020learning}, and adaptive mesh refinement \citep{han2022predicting}. 
\citet{hu2023graph} further demonstrated that graph neural networks can learn effective reduced representations for real-world dynamic systems. More recently, M4GN \citep{leim4gn}, a hierarchical mesh-based graph surrogate have been proposed to better capture long-range interactions and improve the accuracy-efficiency tradeoff. While these approaches have achieved impressive predictive accuracy with speedups exceeding two orders of magnitude compared to traditional solvers \citep{beale1985high}, their latent representations remain largely opaque, hindering deployment in safety-critical applications.

\subsection{Sparse Autoencoders for Mechanistic Interpretability}
Sparse autoencoders (SAEs) are increasingly used for mechanistic interpretability by learning overcomplete, sparse feature dictionaries. \citet{Cunningham2023} show that SAEs trained on language model residual streams can recover highly interpretable features, and \citet{Gao2024} extend this to larger LLMs, identifying scaling laws and quantitative feature-quality metrics. Variants such as gated, k-sparse, L0-regularized, mutual-regularized, and switch SAEs have further improved performance \citep{rajamanoharan2024improving,makhzani2013k,rajamanoharan2024jumping,Marks2024,Mudide2024}, and pretrained SAEs of LLMs are available through Gemma-Scope \citep{lieberum2024gemma}. In vision, SAEs have been used to align concepts across models and enable causal interventions on learned features \citep{Thasarathan2025,Stevens2025}. These advances build on broader work in disentangled representation learning \citep{higgins2017beta,chen2018isolating,locatello2019challenging}. However, prior SAE research has focused mainly on language and vision, and, to the best of our knowledge, has not been applied to physics-based surrogate models to provide disentangled representations suitable for targeted post-hoc intervention, where features must capture continuous, PDE-governed spatiotemporal dynamics rather than discrete semantic concepts.

\subsection{Latent Space Control and Model Steering}
Latent-space steering has emerged as a powerful paradigm for post-hoc control of learned models.
In generative modeling, \citet{shen2020interpreting} showed that GAN latent spaces encode semantically meaningful directions for targeted image editing.
Recent work in language models introduced \emph{activation engineering} techniques that modify internal representations to steer outputs without retraining \citep{subramani2022extracting, li2023inference, turner2023steering, rimsky2024steering, zou2023representation}. For example, \citet{o2025steering} extended these techniques to SAE-derived features for refusal steering, while \citet{kulkarni2025interpretable} introduced concept bottleneck SAEs for interpretable interventions.
However, these methods primarily target static or single-forward-pass scenarios in language and vision, where interventions typically involve scaling, additive offsets, or clamping individual features \citep{o2025steering}.
Such static per-feature interventions suit discrete semantic attributes but do not naturally extend to continuous, time-dependent dynamics.
In dynamical systems, \citet{lusch2018deep} learned linear embeddings of nonlinear dynamics for Koopman-based control, while \citet{brunton2015closed} surveyed closed-loop control strategies for turbulent flows.
Traditional reduced-order modeling techniques such as POD and DMD \citep{brunton2019data,kutz2016dynamic,taira2020modal} provide control-oriented decompositions but operate on state-space observations rather than learned neural representations.
Our work bridges latent-space steering with control of continuous physical dynamics.
We demonstrate that SAE-based steering transfers to time-dependent CFD surrogates when intervention design respects dynamical structure: rather than static per-feature edits, we identify oscillatory pairs via Hilbert analysis and apply temporally coherent, phase-aware rotations that preserve amplitude-phase coupling, enabling real-time phase synchronization without retraining.


\section{Method}
\label{sec:method}
This section describes an end-to-end framework for correcting phase drift in frozen graph-based CFD surrogates through latent-space rotations. The pipeline is representation-agnostic: the same procedure applies identically whether the surrogate embeddings are transformed by a sparse autoencoder (SAE), projected onto principal components (PCA), or left in their raw form. The six subsections below follow the order of the algorithm: formulate the frozen-surrogate setting (Section~\ref{sec:formulation}), identify oscillatory latent pairs (Section~\ref{sec:pairs}), compress each pair into a low-rank spatial mode representation (Section~\ref{sec:svd}), parameterize a smooth time-varying phase offset (Section~\ref{sec:phase_param}), apply the correction via coefficient rotation and roll out through the frozen surrogate (Section~\ref{sec:rotation}), and optimize the steering parameters against available observations (Section~\ref{sec:objective}).

\subsection{Problem Formulation and Frozen-Surrogate Setting}
\label{sec:formulation}
 
\paragraph{Pretrained surrogate.}
We consider a MeshGraphNet (MGN)~\citep{pfaff2020learning} trained on unsteady CFD simulations. For a graph snapshot $G_t = (\mathcal{N}, \mathcal{E})$ at time~$t$, the MGN's encoder--process--decoder pipeline produces a processed node embedding $\mathbf{h}_{t, n}^L \in \mathbb{R}^{d_{\mathrm{emb}}}$ at each node $n \in \mathcal{N}$ after $L$ message-passing iterations. A decoder MLP $f_{dec}$ maps these embeddings to the predicted next-step state: $\hat{\mathbf{x}}_{t+1, n} = f_{dec}(\mathbf{h}_{t, n}^L)$. The surrogate is trained end-to-end with a mean-squared-error loss over snapshots from unsteady CFD simulations:
\begin{equation}
\label{eq:mgn_loss}
\mathcal{L}_{\mathrm{MGN}} = \frac{1}{|\mathcal{N}|} \sum_{n \in \mathcal{N}} \left\| \hat{\mathbf{x}}_{t+1, n} - \mathbf{x}_{t+1, n} \right\|_2^2.
\end{equation}
In the cylinder-flow setting considered here, the predicted state $\hat{\mathbf{x}}_{t+1,n} \in \mathbb{R}^{d}$ consists of the velocity components $(u_x, u_y)$, so $d = 2$.  Collecting these predictions over all
nodes and time steps yields the velocity field used in the steering objective; we denote by $\mathbf{U}^{\mathrm{steer}}$ and $\mathbf{U}^{\mathrm{target}}$ the steered and target velocity sequences, respectively, and are described in detail later.

 
\paragraph{Representation map.}
Let $g\colon \mathbb{R}^{d_{\mathrm{in}}} \to \mathbb{R}^{D}$ denote a generic, fixed representation map applied to the frozen node embeddings. We consider three instantiations:
\begin{itemize}
    \item \emph{Sparse Autoencoder (SAE).} A single-hidden-layer autoencoder with expansion factor $\kappa > 1$ and ReLU activation is trained on the collection of frozen embeddings $\{\mathbf{h}_{i,t}^L\}$. The encoder computes $\mathbf{z} = \sigma\bigl((\mathbf{h} - \mathbf{b}_{\mathrm{dec}})\,\mathbf{W}_{\mathrm{enc}} + \mathbf{b}_{\mathrm{enc}}\bigr)$, where the pre-centering by the decoder bias $b_{\mathrm{dec}}$ follows the convention of \citet{Cunningham2023}, ensuring that the encoder operates on residuals relative to the decoder's learned mean. The decoder reconstructs $\hat{\mathbf{h}} = \mathbf{z}\,\mathbf{W}_{\mathrm{dec}} + \mathbf{b}_{\mathrm{dec}}$, where $\mathbf{W}_{\mathrm{enc}} \in \mathbb{R}^{d_{\mathrm{emb}} \times d_{\mathrm{hid}}}$, $\mathbf{W}_{\mathrm{dec}} \in \mathbb{R}^{d_{\mathrm{hid}} \times d_{\mathrm{emb}}}$, and $d_{\mathrm{hid}} = \kappa\, d_{\mathrm{emb}}$. Training minimizes $\mathcal{L}_{\mathrm{SAE}} = \|\hat{\mathbf{h}} - \mathbf{h}\|_2^2 + \lambda \|\mathbf{z}\|_1$, with the decoder rows renormalized to unit~$\ell_2$ norm after each step. After convergence, the SAE parameters are frozen and $g$ is defined by the encoder, yielding $D = d_{\mathrm{hid}}$.
    
    \item \emph{PCA.} The frozen embeddings are projected onto their top $D_{\mathrm{PCA}}$ principal components. These directions are orthogonal and capture maximum variance, but are dense: every component is a linear combination of all $d_{\mathrm{emb}}$ embedding dimensions. Here $D = D_{\mathrm{PCA}}$.
    
    \item \emph{Identity (raw embedding).} The embedding is left unmodified, so $g$ is the identity map and $D = d_{\mathrm{emb}}$.
\end{itemize}
In all three cases, the inverse map $g^{-1}$ (SAE decoder, inverse PCA projection, or identity) returns modified representations to the MGN embedding space.
 
\paragraph{Horizon and target.}
At deployment, we operate on a finite horizon of $H{+}1$ time steps extracted from the surrogate rollout. Let $\mathbf{X} \in \mathbb{R}^{(H+1) \times N \times D}$ denote the representation-space activations over this horizon, where $N = |\mathcal{N}|$ is the number of mesh nodes and $D$ is the representation dimensionality under map~$g$. Operating on a horizon rather than a single time step is essential for three reasons: (i)~oscillation identification requires temporal context to estimate phase and frequency; (ii)~smoothness regularization of the phase trajectory requires multiple samples; and (iii)~the loss terms used for optimization (velocity matching, temporal derivative alignment) are defined over time differences. 
We denote the desired phase lead or lag by an integer shift
$L_{\mathrm{target}} \in \mathbb{Z}$ (in frames).  The target velocity sequence $\mathbf{U}^{\mathrm{target}} \in \mathbb{R}^{(H+1) \times N \times d}$ is constructed by time-shifting the surrogate's predicted velocity field
$\hat{\mathbf{x}}_{t,n}$ by $L_{\mathrm{target}}$ frames.
 
Note that the dynamics of a cylinder wake are typically categorized into distinct regimes (e.g., near-equilibrium linear dynamics, transient dynamics following a Hopf bifurcation, and periodic limit-cycle dynamics~\citep{chen2012variants}), each exhibiting fundamentally different behavior. The horizon start~$t_0$ should be chosen after the flow has settled into the periodic limit-cycle regime, where vortex shedding is steady and periodic and phase control is applicable.

\subsection{Identification of Oscillatory Pairs}
\label{sec:pairs}
 
The representation-space activations $\mathbf{X}$ contain features spanning a wide range of flow behaviors, but only a subset participate meaningfully in the dominant vortex-shedding oscillation that gives rise to phase drift. This subsection describes a principled procedure for isolating \emph{oscillatory pairs}: pairs of features that jointly form a sine--cosine-like representation of a single underlying periodic mode.
 
\paragraph{Node-averaged time series.}
For each feature $f \in \{1, \ldots, D\}$, we form a node-averaged time series over the horizon:
\begin{equation}
\label{eq:node_avg}
\bar{x}_f(t) = \frac{1}{N} \sum_{n=1}^{N} X_{t,n,f}, \qquad t = t_0, \ldots, t_0 + H.
\end{equation}
Node averaging removes spatially local fluctuations and exposes the globally coherent oscillatory structure of each feature.
 
\paragraph{Hilbert transform and instantaneous phase.}
For each node-averaged time series $\bar{x}_f(t)$, we form its analytic signal:
\begin{equation}
\label{eq:analytic}
\tilde{x}_f(t) = \bar{x}_f(t) + \mathrm{i}\,\mathcal{H}\{\bar{x}_f(t)\},
\end{equation}
where $\mathcal{H}\{\cdot\}$ denotes the Hilbert transform and $\mathrm{i} = \sqrt{-1}$ is the imaginary unit. The instantaneous phase is then
\begin{equation}
\label{eq:inst_phase}
\theta_f(t) = \arg\, \tilde{x}_f(t) = \arctan \frac{\mathcal{H}\{\bar{x}_f(t)\}}{\bar{x}_f(t)}.
\end{equation}
A robust frequency proxy for each feature is obtained from the median phase increment: $\hat{\omega}_f = \mathrm{med}\, |\Delta \theta_f|$.
 
\paragraph{Filtering criteria.}
Two features $(i, j)$ are retained as a candidate oscillatory pair if they satisfy three conditions:
\begin{enumerate}
    \item \emph{Sufficient temporal amplitude.} Both features must exhibit oscillation amplitudes whose $z$-scores exceed a threshold, ensuring that the oscillations are well-resolved above noise.
    \item \emph{Frequency similarity.} The features must oscillate at approximately the same frequency: $|\hat{\omega}_i - \hat{\omega}_j| < \epsilon_\omega$, which avoids pairing features that encode disparate time scales.
    \item \emph{Near-quadrature phase relationship.} The mean phase difference must satisfy $\theta_i(t) - \theta_j(t) \approx \pi/2$, providing a sine--cosine-like basis in which a rotation implements a pure time shift.
\end{enumerate}
 
\paragraph{Ranking.}
After the three hard filters above, multiple candidate pairs may remain. To prioritize the most reliable and physically impactful pairs, we rank them by four complementary metrics:
\begin{itemize}
    \item \emph{Phase coherence.} Stability of the phase difference $\theta_i(t) - \theta_j(t)$ over the horizon, computed as
    $\mathrm{coh}(i,j)
  = \left|\frac{1}{H+1}\sum_{t=t_0}^{t_0+H}
    e^{\,\mathrm{i}(\theta_i(t)-\theta_j(t))}\right|$.
    Values near~1 indicate a temporally consistent sine--cosine relationship.
    
    \item \emph{Amplitude strength.} Average oscillation energy of each feature, measured by the Hilbert envelope or variance of $\bar{x}_f(t)$. Stronger oscillations yield more impactful steering.
    
    \item \emph{Decoder strength.} Contribution of each feature to the surrogate reconstruction, measured by the $\ell_2$ norm of the corresponding decoder column (for SAE) or principal-component loading (for PCA). Features with negligible decoder weights have limited physical influence and are down-ranked.
    
    \item \emph{Spatial footprint coherence.} For each feature $f$ in a candidate pair, we define a per-node energy map
    \begin{equation}
    \label{eq:energy_map}
    E_f(n) = \frac{1}{H+1} \sum_{t=t_0}^{t_0+H} (X_{t,n,f})^2,
    \end{equation}
    which measures how strongly feature~$f$ activates at each mesh node over the horizon. Pairs whose energy maps are spatially co-localized and overlap with physically meaningful flow regions (e.g., shear layers, wake vortices) are prioritized. Co-localization is quantified by the normalized inner product $\langle E_i, E_j \rangle / (\|E_i\|\, \|E_j\|)$.
\end{itemize}
By combining these four metrics, we rank the candidate pool and select the top~$P$ oscillatory pairs $\{(i_k, j_k)\}_{k=1}^P$ to be steered. 
 

\subsection{Low-Rank Spatial Mode Decomposition}
\label{sec:svd}
 
Each selected feature is a high-dimensional spatiotemporal field defined on all $N$ mesh nodes over the horizon. To make phase manipulation computationally tractable and numerically stable, we compress each feature into a low-rank representation via singular value decomposition (SVD), analogous to Proper Orthogonal Decomposition (POD) of velocity fields in classical fluid mechanics~\citep{chatterjee2000introduction, taira2017modal}.
 
For each feature $f$ in a selected pair, we assemble its space--time matrix on the horizon $\mathbf{X}^f \in \mathbb{R}^{(H+1) \times N}$. Subtracting the temporal mean $\boldsymbol{\mu}_f = \frac{1}{H+1} \sum_t \mathbf{X}^f_t \in \mathbb{R}^N$ yields $\mathbf{Z}^f = \mathbf{X}^f - \boldsymbol{\mu}_f$, which isolates the oscillatory content from the time-invariant component. We then compute the SVD $\mathbf{Z}^f = \mathbf{U}_f \boldsymbol{\Sigma}_f \mathbf{V}_f^\top$ and truncate to rank~$r$, obtaining spatial modes
\begin{equation}
\label{eq:spatial_modes}
\boldsymbol{\Phi}_f = \mathbf{V}_f[:, 1{:}r] \in \mathbb{R}^{N \times r}
\end{equation}
and associated time-dependent coefficients
\begin{equation}
\label{eq:time_coeffs}
\mathbf{C}^f(t) = (\mathbf{X}^f_t - \boldsymbol{\mu}_f)\, \boldsymbol{\Phi}_f \in \mathbb{R}^r,
\end{equation}
so that each feature snapshot is approximated as $\mathbf{X}^f_t \approx \mathbf{C}^f(t)\, \boldsymbol{\Phi}_f^\top + \boldsymbol{\mu}_f$. 
Throughout, superscript feature indices (e.g., $\mathbf{C}^{f}$,
$\mathbf{X}^{f}_{t}$) denote activation values \emph{of} feature~$f$,
while subscript feature indices (e.g., $\boldsymbol{\Phi}_{f}$,
$\boldsymbol{\mu}_{f}$) label spatial structures
\emph{associated with} feature~$f$.
The truncation rank~$r$ (e.g., 6--12) is chosen to retain coherent oscillatory energy while discarding noise; in practice, a fixed small~$r$ works well for vortex-shedding horizons.
 
Working in this low-dimensional coefficient space is the key enabler for the phase manipulation that follows. Rotating $r$-dimensional coefficient vectors rather than $N$-dimensional node fields avoids direct intervention in the high-dimensional mesh data, reduces the number of degrees of freedom affected by the correction, and improves numerical conditioning.

\subsection{Time-Varying Phase Parameterization}
\label{sec:phase_param}
 
Even in nominally periodic flows such as vortex shedding, the phase mismatch between the surrogate and the target is not a single constant: small discrepancies in shedding frequency, transient fluctuations, and surrogate prediction bias cause the phase error to drift over time. A fixed phase offset is therefore insufficient for long horizons; the correction must itself evolve smoothly. We parameterize the phase offset for each selected pair~$k$ with features $(i_k, j_k)$ as a low-dimensional, time-varying function:
\begin{equation}
\label{eq:phase_param}
\Delta \phi_k(t) = a_k\, t + b_k + (\mathbf{B}\, \mathbf{w}_k)_t,
\end{equation}
where $a_k, b_k \in \mathbb{R}$ are a learnable slope and offset, $\mathbf{w}_k \in \mathbb{R}^{K_{\mathrm{basis}}}$ are basis weights, and $\mathbf{B} \in \mathbb{R}^{(H+1)\times K_{\mathrm{basis}}}$ is a fixed
low-frequency cosine dictionary with entries $B_{t,m} = \cos\bigl(\tfrac{2\pi m t}{H+1}\bigr)$,
$m = 1, \dots, K_{\mathrm{basis}}$, and unit column normalization. The $m = 0$ (constant) mode is excluded because its effect is already captured by the offset~$b_k$. Each component serves a distinct purpose:
\begin{itemize}
    \item The linear term $a_k t + b_k$ captures persistent frequency bias and global phase offset, which are the dominant sources of drift in deployment.
    \item The cosine basis $\mathbf{B}\,\mathbf{w}_k$ provides smooth, bandwidth-limited adjustments that accommodate slow nonlinear drift without overfitting frame-to-frame noise.
\end{itemize}
Using a fixed $\mathbf{B}$ tied only to the horizon length keeps the optimization low-dimensional and well-conditioned: only $K_{\mathrm{basis}} + 2$ scalars are learned per pair, independent of~$N$. A small value of $K_{\mathrm{basis}}$ (e.g., 4--6) is sufficient in practice to represent smooth phase variations over typical control horizons.

\subsection{Coefficient Rotation, Inverse Mapping, and Rollout}
\label{sec:rotation}
 
\paragraph{Pairwise rotation.}
For each selected oscillatory pair $(i_k, j_k)$, we apply a time-varying
rotation to their SVD coefficient vectors at each time step.  Because
$\mathbf{C}^{i_k}(t),\, \mathbf{C}^{j_k}(t) \in \mathbb{R}^{r}$, the rotation is applied independently
to each SVD-mode index $m \in \{1,\dots,r\}$:
\begin{equation}\label{eq:rotation}
\begin{pmatrix}
  C'^{\,i_k}_{m}(t) \\[4pt]
  C'^{\,j_k}_{m}(t)
\end{pmatrix}
=
\begin{pmatrix}
  \cos \Delta\phi_k(t) & -\sin \Delta\phi_k(t) \\
  \sin \Delta\phi_k(t) & \phantom{-}\cos \Delta\phi_k(t)
\end{pmatrix}
\begin{pmatrix}
  C^{i_k}_{m}(t) \\[4pt]
  C^{j_k}_{m}(t)
\end{pmatrix},
\quad m = 1, \dots, r.
\end{equation}
The rotated components are reassembled into the full coefficient vectors
$\mathbf{C}'^{\,i_k}(t),\, \mathbf{C}'^{\,j_k}(t) \in \mathbb{R}^{r}$,
which are then used in the reconstruction below.

Because the two features form a near-quadrature pair, this rotation in the $(\mathbf{C}^{i_k}, \mathbf{C}^{j_k})$ plane is equivalent to a phase (time) shift of the underlying oscillation: it advances or retards the periodic mode without altering its amplitude or spatial structure. This is the central distinction from static steering methods (scaling, additive perturbation, clamping), which modify features independently and cannot preserve the coupled amplitude--phase relationship that defines a coherent oscillation.
 
\paragraph{Reconstruction and inverse mapping.}
The rotated feature fields are reconstructed from the modified coefficients:
\begin{equation}
\label{eq:reconstruct}
\mathbf{X}'^{i_k}_t \approx \mathbf{C}'^{i_k}(t)\, \boldsymbol{\Phi}_{i_k}^\top + \boldsymbol{\mu}_{i_k}, \qquad
\mathbf{X}'^{j_k}_t \approx \mathbf{C}'^{j_k}(t)\, \boldsymbol{\Phi}_{j_k}^\top + \boldsymbol{\mu}_{j_k}.
\end{equation}
All features not belonging to a selected pair are left unchanged, yielding the steered representation tensor $\mathbf{X}' \in \mathbb{R}^{(H+1) \times N \times D}$. The inverse representation map~$g^{-1}$ (SAE decoder, inverse PCA projection, or identity) then returns $\mathbf{X}'$ to the MGN embedding space, and the frozen MGN decoder produces a steered velocity sequence $\mathbf{U}^\mathrm{steer} \in \mathbb{R}^{(H+1) \times N \times d}$.

\subsection{Objective and Optimization}
\label{sec:objective}
 
\paragraph{Loss function.}
The steering parameters $\{a_k, b_k, \mathbf{w}_k\}_{k=1}^P$ are optimized by minimizing a composite loss that combines state-based alignment, phase alignment, and regularization:
\begin{equation}
\label{eq:total_loss}
\mathcal{L} = \lambda_{\mathrm{vel}}\, \mathcal{L}_{\mathrm{vel}} + \lambda_{\mathrm{dv}}\, \mathcal{L}_{\mathrm{dv}} + \lambda_{\mathrm{phase}}\, \mathcal{L}_{\mathrm{curv}} + \lambda_{\mathrm{mag}}\, \mathcal{L}_{\mathrm{mag}}.
\end{equation}
The individual terms are defined as follows.
\begin{itemize}
    \item \emph{Velocity alignment} matches the steered velocities to the target:
    \begin{equation}
    \label{eq:loss_vel}
    \mathcal{L}_{\mathrm{vel}} = \frac{1}{(H{+}1)\, N} \sum_{t=t_0}^{t_0+H} \sum_{n=1}^{N} \bigl\| U^{\mathrm{steer}}_{t,n} - U^{\mathrm{target}}_{t,n} \bigr\|^2,
    \end{equation}
    where the sums range over all $H{+}1$ time steps in the steering horizon and all $N$ mesh nodes.
    \item \emph{Temporal derivative alignment} matches discrete temporal derivatives, discouraging jitter and promoting dynamically consistent rollouts:
    \begin{equation}
    \label{eq:loss_dv}
    \mathcal{L}_{\mathrm{dv}} = \frac{1}{H\, N} \sum_{t=t_0}^{t_0+H-1} \sum_{n=1}^{N} \bigl\| (U^{\mathrm{steer}}_{t+1,n} - U^{\mathrm{steer}}_{t,n}) 
     - (U^{\mathrm{target}}_{t+1,n} - U^{\mathrm{target}}_{t,n}) \bigr\|^2.
    \end{equation}
     
     
    \item \emph{Curvature regularization} enforces smoothness of the learned phase trajectories:
    \begin{equation}
    \label{eq:loss_curv}
    \mathcal{L}_{\mathrm{curv}} = \frac{1}{P\,(H{-}1)} \sum_{k=1}^{P} \sum_{t=t_0}^{t_0+H-2} \bigl(\Delta\phi_k(t) - 2\,\Delta\phi_k(t{+}1) 
     + \Delta\phi_k(t{+}2)\bigr)^2.
    \end{equation}
     
    \item \emph{Magnitude regularization} prevents the steered embeddings from drifting far from their unsteered counterparts in the MGN  embedding space, isolating the effect of steering from any reconstruction error introduced by the representation map:
    \begin{equation}
    \mathcal{L}_{\mathrm{mag}} = \frac{1}{(H{+}1)\, N\, d_{\mathrm{emb}}}
    \sum_{t=t_0}^{t_0+H} \sum_{n=1}^{N}
    \bigl\| g^{-1}(X'_{t,n}) - g^{-1}(X_{t,n}) \bigr\|^2,
    \end{equation}
    where $X'_{t,n}$ and $X_{t,n}$ are the steered and unsteered representation vectors at node $n$ and time $t$, and $g^{-1}$  maps both back to the $d_{\mathrm{in}}$-dimensional MGN embedding  space. 
\end{itemize}
\paragraph{Evaluated setting.}
In the experiments reported in this paper, full flow-field data from a high-fidelity simulation are available. The target sequence is constructed by shifting the unsteered surrogate prediction by $L_{\mathrm{target}}$ frames. All state-based loss terms ($\mathcal{L}_{\mathrm{vel}}$, $\mathcal{L}_{\mathrm{dv}}$) are computed over the full node set, and both regularization terms ($\mathcal{L}_{\mathrm{curv}}$, $\mathcal{L}_{\mathrm{mag}}$) are active. Only the low-dimensional steering parameters $\{a_k, b_k, \mathbf{w}_k\}$ are updated; the surrogate, the SAE, and the spatial modes $\boldsymbol{\Phi}_f$ all remain frozen.
 

\section{Experimental Setup}
\label{sec:experiments}

This section specifies how SAE, PCA, and raw-embedding representations are compared under the same rotation-based steering task, and how static intervention baselines are constructed. All design choices, including dataset, surrogate architecture, SAE training, steering pipeline, hyperparameter sweep, and metrics, are documented to enable reproducibility.

\subsection{Dataset and Base Surrogate}
\label{sec:dataset}

We use the CylinderFlow dataset~\citep{pfaff2020learning}, which comprises simulations of transient incompressible flow around a cylinder with varying diameters and positions on a fixed two-dimensional Eulerian mesh. The dataset contains 1{,}000 training, 100 validation, and 100 test simulations, each spanning 600 time steps. Node types distinguish among fluid nodes, wall nodes, and inflow/outflow boundary nodes; the inlet boundary condition is a prescribed parabolic velocity profile.

The base surrogate is a MeshGraphNet (MGN)~\citep{pfaff2020learning} trained with the same hyperparameter configuration as described in the original paper: nine message-passing iterations, a latent dimension of 128 for both node and edge features, residual MLPs with two hidden layers and layer normalization in each update module. After training, the MGN weights are frozen and are not modified at any point during SAE training, steering, or evaluation.

\subsection{SAE Training}
\label{sec:sae_training}

The sparse autoencoder is trained post-hoc on the frozen node embeddings $\{\mathbf{h}_{i,t}^L\}$ produced by the trained MGN. We use an expansion factor $\kappa = 8$, yielding a hidden-layer width of $d_{\mathrm{hid}} = 8 \times 128 = 1{,}024$. The sparsity coefficient is set to $\lambda = 3 \times 10^{-4}$, and training is performed with a mini-batch size of 128 using the Adam optimizer with a learning rate of $1 \times 10^{-3}$. Training proceeds until the reconstruction loss on a held-out validation set stops decreasing. After convergence, the SAE parameters are frozen. 

\subsection{Representations Compared}
\label{sec:representations}

We evaluate the steering framework across three representation maps $g$ (defined in Section~\ref{sec:formulation}), which differ only in how the frozen MGN embeddings are transformed before steering:
\begin{enumerate}
    \item \textbf{Sparse Autoencoder (SAE):} An overcomplete, sparse representation with expansion factor $\kappa = 8$ ($D = 1{,}024$), trained as described in Section~\ref{sec:sae_training}. The SAE produces a disentangled dictionary in which ${\sim}87.6\%$ of activations are exactly zero at any given time step.
    
    \item \textbf{PCA:} A dense, decorrelated representation obtained by projecting the 128-dimensional MGN embeddings onto their principal components. PCA directions are orthogonal and capture maximum variance but are not sparse: every component is a linear combination of all 128 embedding dimensions.
    
    \item \textbf{Raw MGN embedding:} The unprocessed 128-dimensional node embeddings produced by the surrogate's encoder--process stage ($g$ is the identity map). These embeddings are neither sparse nor decorrelated.
\end{enumerate}
Crucially, the steering pipeline of Section~\ref{sec:pairs}--\ref{sec:objective} is applied identically in all three cases: oscillatory pairs are identified, SVD-decomposed, and rotated using the same parameterization and the same optimization objective. Only the representation space differs. This controlled design isolates the effect of representation quality on steering performance.

\subsection{Evaluation Protocol}
\label{sec:protocol}

\paragraph{Task specification.}
Phase steering is evaluated with a target shift of $L_{\mathrm{target}} = +8$ frames (approximately one-third of a shedding period) as a representative nontrivial phase offset: it is large enough to require meaningful correction, yet small enough that the target remains within the single-cycle phase-steering regime considered in this proof-of-concept study. All methods share the same frozen MeshGraphNet, the same SVD truncation rank, and the same Adam optimizer.

\paragraph{Implementation details.}
The steering horizon is $H = 120$ frames, starting at $t_0 = 140$ after the flow has settled into the periodic limit-cycle regime.
The SVD truncation rank is $r = 8$ for all representations, and the phase parameterization uses $K_\mathrm{basis} = 6$ cosine basis functions. For each representation, we sweep over the number of oscillatory pairs $P \in \{4, 5, 6, 7, 8\}$ and the magnitude regularization weight $\lambda_{\mathrm{mag}} \in \{10^{-4}, 10^{-3}, 5 \times 10^{-3}\}$. We focus the sweep on these two hyperparameters because they most directly govern the trade-off between steering aggressiveness and surrogate consistency: $P$ controls how many oscillatory modes are corrected, and $\lambda_{\mathrm{mag}}$ controls how far the steered embeddings may deviate from the original. For each representation, the configuration that maximizes $\mathrm{frac\%}(v_x)$ is selected and reported.


\paragraph{A note on model selection.}
In the current experiments, configuration selection and final evaluation are both performed on the same test trajectory. This design is appropriate for a controlled proof-of-concept whose primary aim is to compare representation quality under identical conditions, but it does not constitute a fully cross-validated benchmark. A production deployment would select $(P, \lambda_{\mathrm{mag}})$ on a held-out validation trajectory or a separate steering window and evaluate once on the test trajectory. We report results from the full sweep (including the Pareto analysis in Section~\ref{sec:sensitivity}) to provide transparency into performance sensitivity across configurations.

\subsection{Baselines and Metrics}
\label{sec:baselines}

\paragraph{Baselines.}
We organize all comparison methods into two groups:
\begin{itemize}
    \item \emph{Rotation-based steering (ours).} The phase-rotation pipeline of Section~\ref{sec:pairs}--\ref{sec:objective}, applied in each of the three representation spaces: SAE, PCA, and raw MGN embedding. This group tests which representation best supports the same physically guided intervention.
    
    \item \emph{Static interventions.} Three standard per-feature manipulation strategies commonly used for SAE-based steering in language and vision models, all applied in the SAE latent space. Unlike rotation-based steering, which operates on feature pairs requiring matched frequencies and near-quadrature coupling, static interventions manipulate features independently. We select the top 10 individual features ranked by the product of oscillation amplitude, decoder gain, and spectral concentration—the same scoring used to build the candidate pool for pair selection in Section~\ref{sec:pairs}, but without the quadrature-coupling constraint:
    \begin{enumerate}
        \item \textbf{Scale:} Multiply each selected feature's activation by an optimized scalar factor.
        \item \textbf{Additive:} Add an optimized constant offset to each selected feature's activation.
        \item \textbf{Clamp:} Fix each selected feature's activation to an optimized constant value across all time steps.
    \end{enumerate}
    This group tests whether standard static interventions, which are effective for editing relatively stable semantic attributes, transfer to a time-dependent dynamical setting.
\end{itemize}

\paragraph{Metrics.}
We evaluate steering performance using four complementary metrics:
\begin{itemize}
    \item $\mathrm{frac\%}$: Measures what percentage of the original-to-target MSE gap is closed by steering, quantifying the overall corrective effect:
    \begin{equation}
    \label{eq:frac}
    \mathrm{frac\%} = \left(1 - \frac{\mathrm{MSE}(v^{\mathrm{steer}},\, v^{\mathrm{target}})}{\mathrm{MSE}(v^{\mathrm{orig}},\, v^{\mathrm{target}})}\right) \times 100.
    \end{equation}
    A positive value indicates that steering moves the prediction closer to the target; a negative value indicates degradation.
    
    \item $\mathrm{ROI\%}$: The same definition as $\mathrm{frac\%}$ but restricted to the downstream vortex-shedding region, which isolates the improvement in the flow region where phase steering is expected to have the greatest impact.
    
    \item $\mathrm{nRMSE}$: Normalized root-mean-square error of the steered field relative to the target, divided by the target RMS. A value below~1 indicates that the steered prediction is closer to the target than the unsteered original, which is a necessary condition for the steering to be considered genuinely corrective.
    
    \item $\mathrm{Corr}$: Pearson correlation between the steered and target velocity fields, measuring spatial pattern agreement independently of amplitude.
\end{itemize}

\section{Results and Evaluation}
\label{sec:results}
The subsections below present overall steering performance across representations (Section~\ref{sec:main_result}), examine where corrections localize spatially (Section~\ref{sec:spatial}), analyze the sparsity and disentanglement properties underlying SAE's advantage (Section~\ref{sec:sae_features}), characterize the oscillatory pairs selected for steering (Section~\ref{sec:pair_structure}), ablate against static interventions (Section~\ref{sec:static_fail}), assess hyperparameter sensitivity (Section~\ref{sec:sensitivity}), and synthesize the findings (Section~\ref{sec:synthesis}).

\subsection{Overall Steering Performance}
\label{sec:main_result}

Table~\ref{tab:main} answers the paper's central question. Under the same rotation-based steering pipeline, SAE achieves a $\mathrm{frac\%}(v_x)$ of $+26.1\%$ and $\mathrm{ROI\%}(v_x)$ of $+35.0\%$, outperforming PCA ($+16.0\%$/$+21.0\%$) by roughly 10/14 percentage points and Raw ($+4.1\%$/$+7.5\%$) by more than 20 percentage points. SAE is also the only representation to attain $\mathrm{nRMSE}(v_x) < 1$, meaning the steered field is closer to the target than the unsteered original---a necessary condition for the steering to be considered genuinely corrective. The $\mathrm{Corr}(v_x v_y)$ metric further separates SAE (0.468) from both PCA (0.367) and Raw (0.359). 
\begin{table}[t]
\centering
\caption{Phase steering performance comparison. \emph{Rotation-based steering} applies the physics-guided rotation pipeline across three embedding spaces (SAE, PCA, Raw). \emph{Static interventions} apply standard per-feature manipulations (Scale, Additive, Clamp) in the SAE latent space. All metrics are computed on the cylinder-flow test trajectory. Best value in each row is \textbf{bolded}.}
\label{tab:main}
\small
\begin{tabular}{l ccc ccc}
\toprule
& \multicolumn{3}{c}{Rotation-based steering (ours)} & \multicolumn{3}{c}{Static interventions (baselines)} \\
\cmidrule(lr){2-4} \cmidrule(lr){5-7}
Metric & SAE & PCA & Raw & Scale & Additive & Clamp \\
\midrule
$\mathrm{frac\%}\,(v_x)\;\uparrow$     & $\mathbf{+26.1\%}$ & $+16.0\%$ & $+4.1\%$  & $-118.6\%$ & $+0.0\%$ & $-494.1\%$ \\
$\mathrm{ROI\%}\,(v_x)\;\uparrow$      & $\mathbf{+35.0\%}$ & $+21.0\%$ & $+7.5\%$  & $-81.4\%$  & $+0.0\%$ & $-149.6\%$ \\
$\mathrm{nRMSE}\,(v_x)\;\downarrow$    & $\mathbf{0.9926}$  & $1.0582$  & $1.1305$  & $1.7067$   & $1.1542$ & $2.8138$   \\
$\mathrm{Corr}\,(v_x v_y)\;\uparrow$   & $\mathbf{0.4681}$  & $0.3666$  & $0.3587$  & $0.1333$   & $0.3384$ & $0.0560$   \\
\bottomrule
\end{tabular}
\end{table}

\begin{figure}[t]
    \centering
    \includegraphics[width=\textwidth]{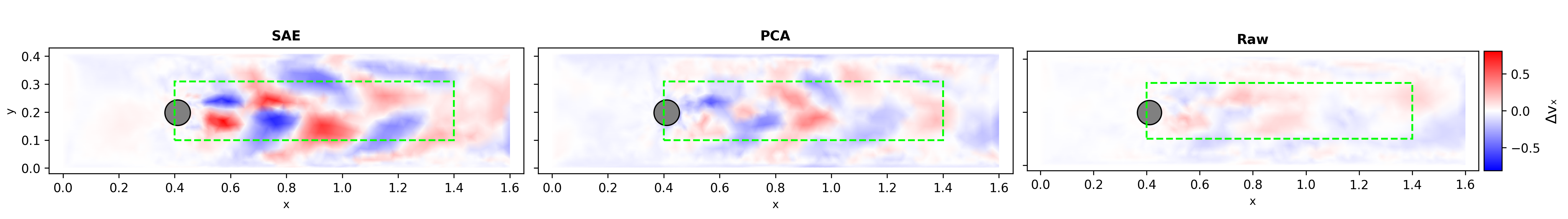}
    \caption{Velocity correction $\Delta v_x = v_x^{\mathrm{steer}} - v_x^{\mathrm{orig}}$ at a selected time frame for the three rotation-based methods. The green dashed rectangle marks the wake ROI. SAE produces the strongest, most spatially coherent correction in the near-wake region.}
    \label{fig:delta_vx}
\end{figure}

Figure~\ref{fig:delta_vx} visualizes the velocity correction $\Delta v_x = v_x^{\mathrm{steer}} - v_x^{\mathrm{orig}}$ at a selected frame for all three rotation-based methods. SAE produces a spatially coherent correction concentrated in the near wake, consistent with a genuine phase advance of the vortex-street pattern. PCA yields a moderate correction with broader spatial spread, while Raw corrections are weak in amplitude and spatially diffuse.

\begin{takeaway}
\textbf{Which representation steers best?}
SAE substantially outperforms PCA and raw embeddings under the same rotation-based steering pipeline, and is the only representation to achieve genuinely corrective predictions. The performance gap is consistent across all four metrics.
\end{takeaway}

\subsection{Spatial Localization of Steering Corrections}
\label{sec:spatial}

\begin{figure}[t]
    \centering
    \includegraphics[width=0.9\textwidth]{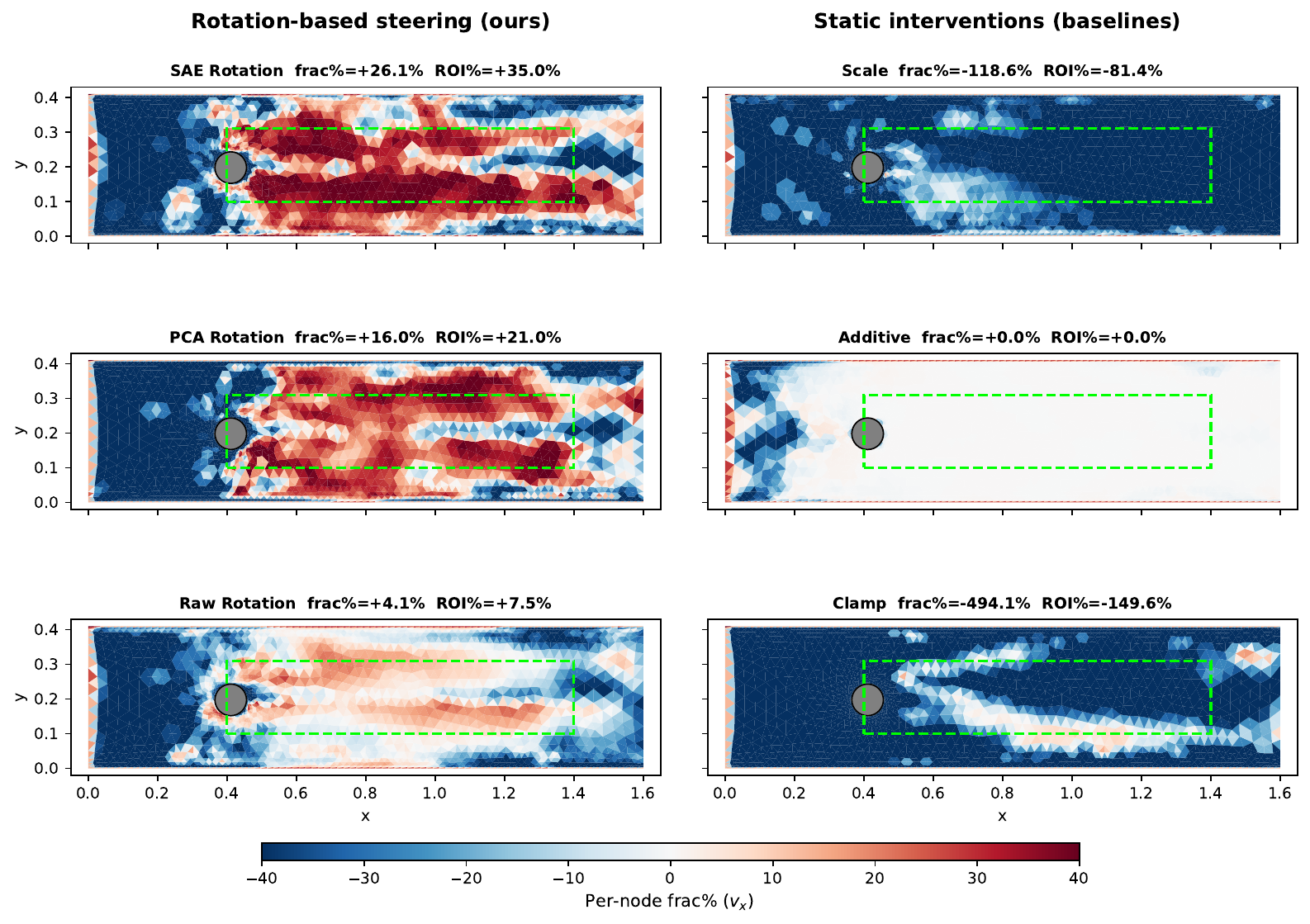}
    \caption{Spatial distribution of per-node $\mathrm{frac\%}(v_x)$. \emph{Left column:} Rotation-based steering across three embedding spaces. \emph{Right column:} Static SAE interventions. Blue indicates improvement; red indicates degradation. Dashed green rectangles mark the wake ROI. Rotation-based steering produces structured, wake-localized improvement, with SAE showing the strongest and most spatially extensive effect. Static interventions produce either no discernible effect (Additive) or widespread degradation (Scale, Clamp).}
    \label{fig:spatial_frac}
\end{figure}

Figure~\ref{fig:spatial_frac} (left column) maps the per-node $\mathrm{frac\%}(v_x)$ across the full domain for each rotation-based method. Two patterns are evident. First, all three methods concentrate improvement downstream of the cylinder in the vortex-shedding region, confirming that the SVD rotation predominantly targets oscillatory modes. Second, SAE produces the most intense and spatially extensive improvement, with individual nodes exceeding $+40\%$ $\mathrm{frac\%}$ in the core wake. PCA achieves moderate improvement concentrated in the near wake but with less spatial extent, while Raw produces only marginal changes.

\begin{figure}[t]
    \centering
    \includegraphics[width=0.8\textwidth]{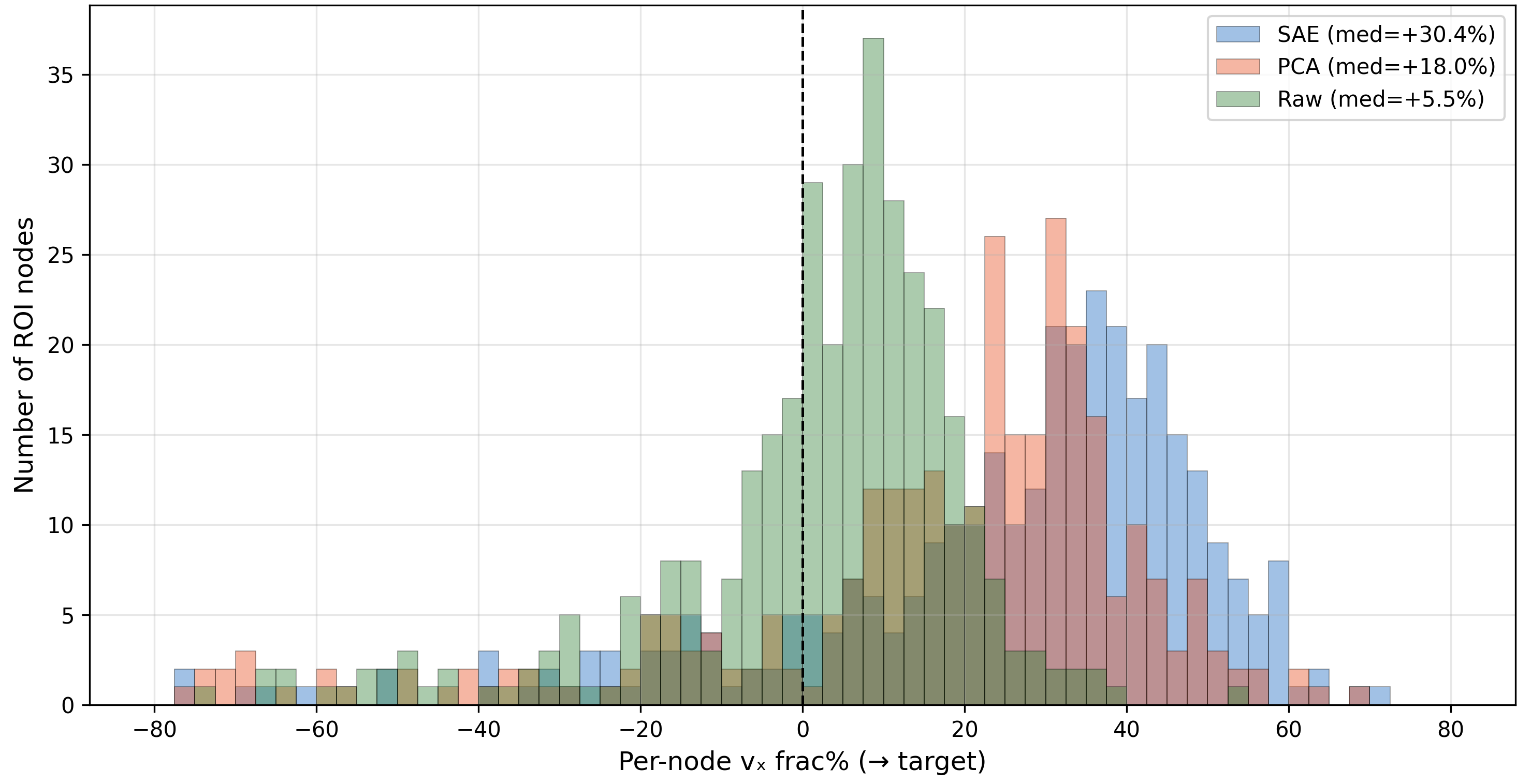}
    \caption{Distribution of per-node $\mathrm{frac\%}(v_x)$ within the wake ROI. SAE shifts the entire distribution rightward, indicating broad-based improvement rather than localized artifacts.}
    \label{fig:roi_hist}
\end{figure}

The ROI histogram (Figure~\ref{fig:roi_hist}) further quantifies the per-node improvement distribution within the wake region ($x \in [0.4, 1.4]$, $y \in [0.10, 0.31]$; 389 nodes). SAE shifts the entire distribution toward positive $\mathrm{frac\%}$, with a median ROI-node improvement of roughly $+30\%$, indicating broad-based correction rather than localized artifacts. Raw barely moves the distribution away from zero. PCA occupies an intermediate position, with a rightward-shifted distribution but a broader tail of degraded nodes than SAE.

\begin{takeaway}
\textbf{Where does steering help?}
All three representations concentrate improvement in the downstream vortex-shedding region, but SAE produces the strongest, most spatially extensive, and most coherent wake-localized correction. 
\end{takeaway}

\subsection{Sparsity and Disentanglement of SAE Features }
\label{sec:sae_features}
The preceding subsections establish that SAE-based steering substantially outperforms PCA and raw-embedding steering, with improvement concentrated in the physically relevant wake region. We now examine the SAE dictionary to identify which properties of the representation account for this advantage.
 
The trained SAE expands the 128-dimensional MGN embedding into 1{,}024 features, of which ${\sim}87.6\%$ are exactly zero at any given time step and node, with a Gini coefficient of 0.863 over temporal variance of the active features. This indicates that oscillatory energy is concentrated in a small subset of the dictionary. In contrast, PCA produces 128 dense components where every direction is a global linear combination of all embedding dimensions, and the raw MGN embedding is neither sparse nor decorrelated.
 
\begin{figure}[t]
    \centering
    \includegraphics[width=0.9\textwidth]{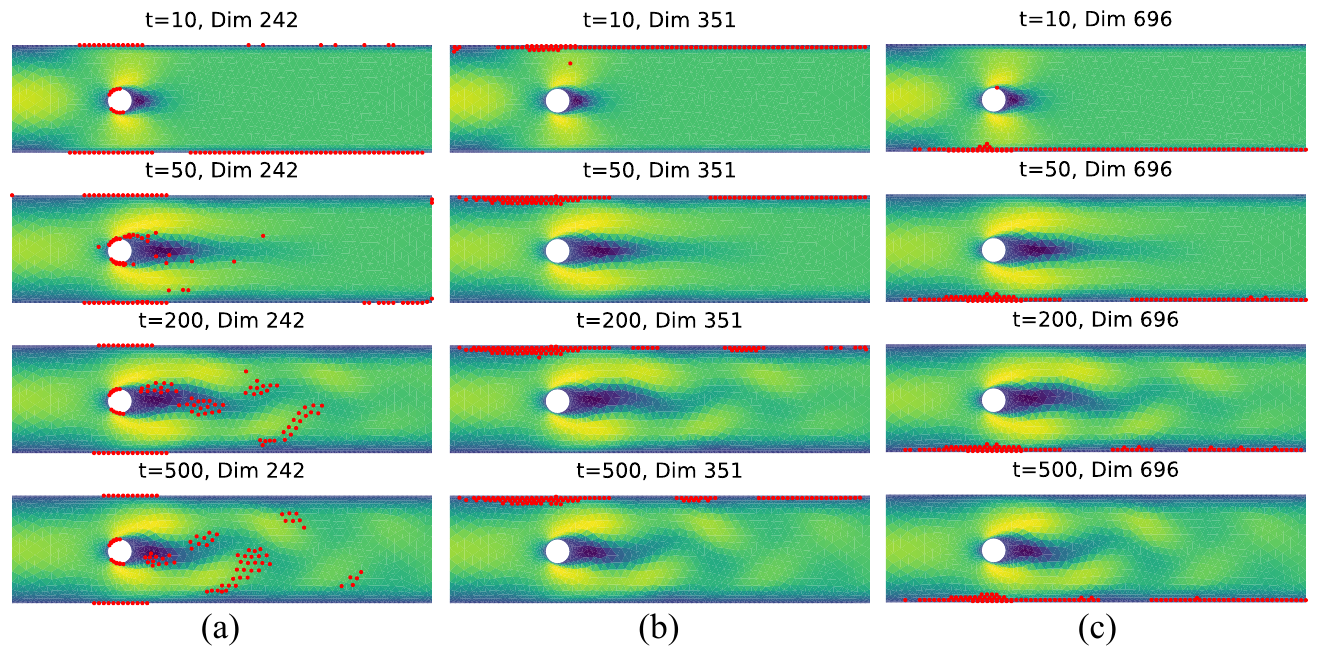}
    \caption{Spatial footprint of three individual salient SAE dimensions selected via the mean-absolute metric. Red markers show the $\eta$ most active nodes ($\eta = 100$) at four representative time steps. Each dimension localizes to a distinct flow feature.}
    \label{fig:sae_dictionary}
\end{figure}

Figure~\ref{fig:sae_dictionary} highlights three representative dimensions from the mean-absolute top representative salient dimensions and visualizes across four snapshots. The clear spatial disjointness of these dimensions confirms
that the SAE dictionary disentangles disparate physical phenomena, underscoring the surrogate’s interpretability and enabling feature-specific diagnostics or control.

This sparsity and spatial disjointness directly explain the steering advantage observed in Section~\ref{sec:main_result}--\ref{sec:spatial}. Because the SAE isolates oscillatory content into a small number of features with localized spatial footprints, the Hilbert-based pair identification (Section~\ref{sec:pairs}) can select pairs that correspond cleanly to the physical vortex-shedding mode. PCA and raw embeddings, lacking this structure, inevitably couple shedding dynamics with unrelated flow physics when steered. A comprehensive quantitative evaluation of the SAE dictionary (including saliency ranking, temporal stability analysis, and comparison against embedding-norm, PCA, and random baselines for vortex-region alignment) can be found in our previous work~\cite{hu2025interpreting}.
 
\subsection{Characterization of Steering-Selected Pairs}
\label{sec:pair_structure}

Having established that the SAE dictionary is sparse and disentangled at the level of individual features, we now examine whether the specific oscillatory pairs (Section~\ref{sec:pairs}) for steering encode physically coherent shedding dynamics---and whether they satisfy the structural prerequisites for rotation-based phase correction.
 
\begin{figure}[t]
    \centering
    \includegraphics[width=1\textwidth]{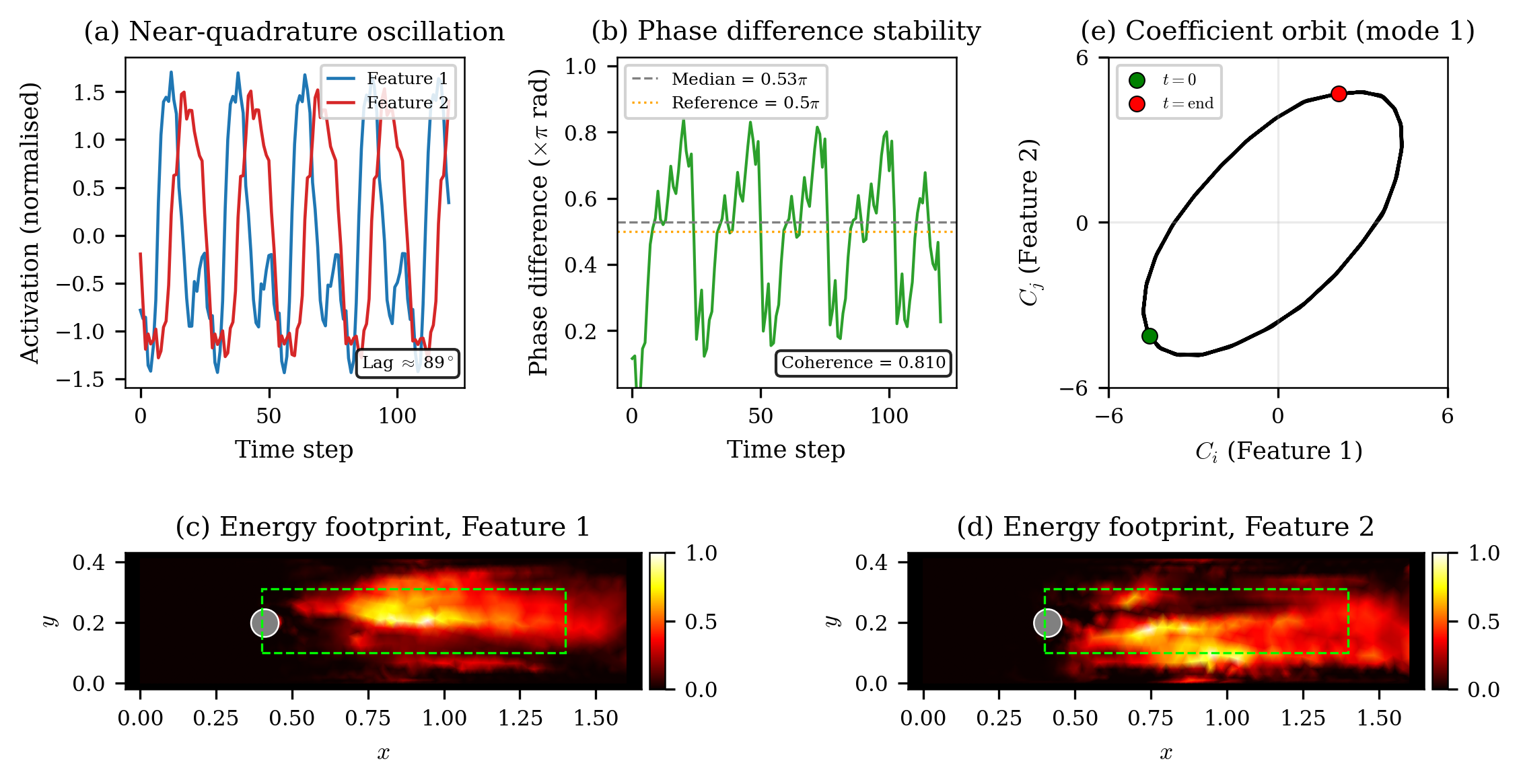}
    \caption{Characterization of oscillatory SAE feature pairs selected for phase-rotation steering within the steering horizon ($H = 120$). (a)~Node-averaged, normalized activation time series of a near-quadrature pair, exhibiting a phase lag of approximately $89^\circ$ at the vortex-shedding frequency. (b)~Instantaneous Hilbert-transform phase difference for the same pair, showing stable fluctuations around the median ($0.53\pi$) with coherence~0.810; the reference line at $0.5\pi$ indicates ideal quadrature. (c,~d)~Spatial energy footprints of a complementary pair, computed as the variance-weighted sum of the leading six SVD spatial modes. The gray circle marks the cylinder; the dashed green rectangle delineates the wake ROI. (e)~Phase-space orbit of leading SVD coefficients for a selected pair, forming a smooth elliptical trajectory characteristic of coupled periodic oscillation. The pairwise rotation advances or retards the trajectory along this ellipse to implement phase correction. Green and red markers indicate the initial and final time steps, respectively.}
    \label{fig:pair_structure}
\end{figure}

Figure~\ref{fig:pair_structure} dissects representative steering-selected pairs across all stages of the selection and rotation pipeline. Panel~(a) overlays the node-averaged, normalized activation time series of a near-quadrature pair: the two features oscillate at the vortex-shedding frequency with a phase lag of approximately $89^\circ$, closely matching the ideal $90^\circ$ required for a sine--cosine basis. Panel~(b) confirms the temporal stability of this relationship via the Hilbert-transform instantaneous phase difference, which fluctuates around a median of $0.53\pi$ with a coherence of 0.810; the reference line at $0.5\pi$ marks ideal quadrature. The deviation from exact $\pi/2$ is small and stable, indicating that the pair reliably encodes a single periodic mode over the steering horizon.
 
Panels~(c--d) display the spatial energy footprints of a complementary pair, computed as the variance-weighted sum of the leading six SVD spatial modes. Feature~1 concentrates its energy in the near-wake region immediately behind the cylinder, while Feature~2 extends further downstream into the far wake. This spatial separation confirms that the SAE dictionary disentangles the wake into localized structures with distinct spatial support, even among features that are jointly selected for steering. The green dashed rectangle delineates the wake region of interest (ROI) used for the $\mathrm{ROI\%}$ metric.
 
Panel~(e) plots the phase-space orbit of the leading SVD coefficients $(C_1^{i_k}, C_1^{j_k})$ for a selected pair, forming a smooth elliptical trajectory characteristic of coupled periodic oscillation. The elliptical geometry is precisely the structure exploited by the pairwise rotation (Section~\ref{sec:rotation}): rotating the coefficient pair by $\Delta\phi_k(t)$ advances or retards the trajectory along this ellipse, implementing a temporal phase shift without distorting the oscillation geometry or altering its amplitude.
 
Together, the dictionary-level evidence (Section~\ref{sec:sae_features}) and the pair-level evidence in this subsection confirm that the steering pipeline operates on physically meaningful coordinates: sparse features that localize to coherent flow structures, paired by quadrature relationships whose elliptical coefficient-space geometry enables clean phase manipulation via rotation.

\begin{takeaway}
\textbf{Are the steered features physically meaningful?}
Yes. The selected pairs oscillate in near-quadrature at the shedding frequency, localize to the wake region, and trace smooth elliptical orbits in coefficient space. 
\end{takeaway}

\subsection{Ablation: Static Interventions}
\label{sec:static_fail}

Table~\ref{tab:main} (right three columns) reports the performance of standard static per-feature interventions applied in the SAE latent space. All three methods fail to improve, and most catastrophically degrade, the surrogate's predictions.

\paragraph{Scale.} Multiplying oscillatory feature activations by an optimized scalar disrupts the amplitude--phase balance: the amplified features overshoot during parts of the cycle and undershoot during others, pushing predictions substantially further from the target ($\mathrm{frac\%} = -118.6\%$).

\paragraph{Additive.} Adding a constant offset to each feature's activation has essentially no effect ($\mathrm{frac\%} = +0.0\%$). This outcome is expected: the SAE decoder absorbs the constant shift into the bias, and the surrogate's autoregressive rollout produces nearly identical dynamics.

\paragraph{Clamp.} Fixing feature activations to a constant value across all time steps destroys temporal coherence entirely, producing catastrophic degradation ($\mathrm{frac\%} = -494.1\%$). The clamped features can no longer track the physical oscillation, and the surrogate's rollout diverges.

Figure~\ref{fig:spatial_frac} (right column) visualizes the spatial distribution of these failures. Scale and Clamp produce widespread degradation (red) across the domain, while Additive shows no discernible spatial pattern, consistent with its near-zero effect. In contrast, the rotation-based methods (left column) produce structured, wake-localized improvement (blue) concentrated in the vortex-shedding region.

These results justify the need for a temporally coherent intervention mechanism rather than direct adoption of the standard SAE steering toolkit from language and vision. In a time-dependent dynamical system, oscillatory features encode both phase and amplitude in a temporally coupled manner; a static scalar intervention cannot disentangle these components. The rotation-based approach succeeds because it operates in the sine--cosine subspace of each oscillatory pair, applying time-varying corrections that preserve amplitude while smoothly adjusting phase.

\begin{takeaway}
\textbf{Do standard static interventions transfer to this setting?}
No. Scaling disrupts amplitude--phase balance, additive offsets have no dynamical effect, and clamping destroys temporal coherence. Effective steering of oscillatory surrogates requires a temporally coherent, structure-preserving intervention, instead of independent per-feature edits.
\end{takeaway}

\subsection{Hyperparameter Sensitivity}
\label{sec:sensitivity}

\begin{figure}[t]
    \centering
    \includegraphics[width=0.7\textwidth]{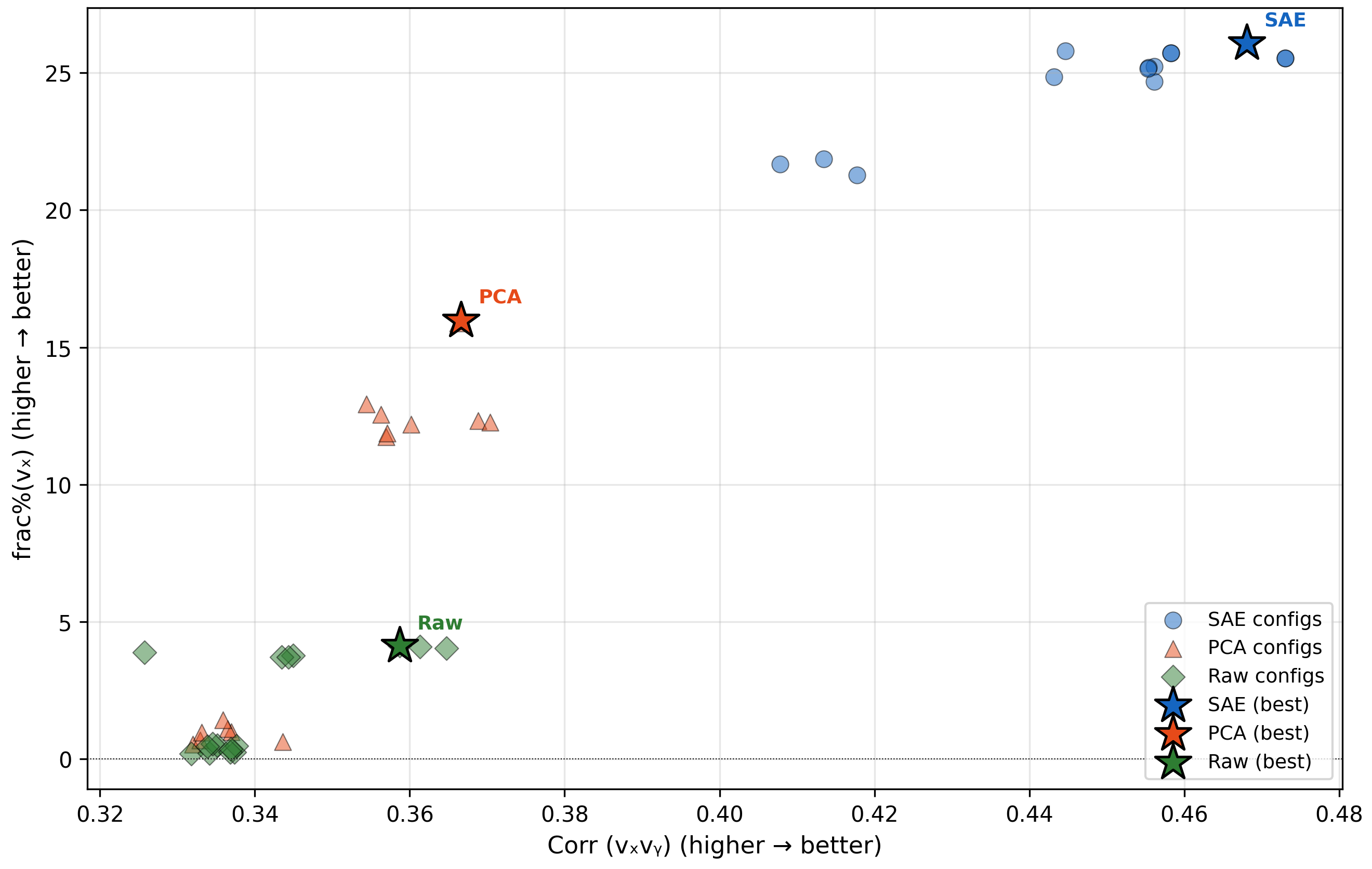}
    \caption{Pareto frontier of $\mathrm{frac\%}(v_x)$ vs.\ $\mathrm{Corr}(v_x v_y)$ across all swept $(P, \lambda_{\mathrm{mag}})$ configurations. Each small marker is one configuration; starred markers denote the best configuration per representation. SAE configurations consistently occupy the upper-right region, dominating both PCA and Raw.}
    \label{fig:pareto}
\end{figure}

Figure~\ref{fig:pareto} plots every swept configuration in the $\mathrm{Corr}(v_x v_y)$--$\mathrm{frac\%}(v_x)$ plane. Three patterns emerge.

First, the SAE point cloud consistently occupies the upper-right quadrant, dominating both PCA and Raw across the entire sweep. The starred markers show the auto-selected best configuration for each representation, all of which sit on or near the Pareto frontier for their respective class.

Second, SAE performance is robust to $\lambda_{\mathrm{mag}}$: across all 15 SAE configurations, $\mathrm{frac\%}$ ranges from $+21.3\%$ to $+26.1\%$, a span of only ${\sim}5$ percentage points. PCA and Raw exhibit similar insensitivity to $\lambda_{\mathrm{mag}}$ but at a lower absolute level.

Third, the number of oscillatory pairs $P$ has a larger effect: SAE performance saturates around $P = 5$--$7$, with diminishing returns thereafter. This saturation is consistent with the expectation that only a small number of SAE feature pairs participate in the dominant shedding mode; additional pairs contribute progressively less oscillatory energy and may introduce spurious coupling.

\begin{takeaway}
\textbf{Is the SAE advantage robust to hyperparameter choices?}
Yes. SAE dominates PCA and raw embeddings across the full $(P, \lambda_{\mathrm{mag}})$ sweep, with performance remaining stable across all tested configurations. 
\end{takeaway}

\subsection{Why SAE Works?}
\label{sec:synthesis}

The results above point to a consistent explanation. The SAE expands the MGN embedding into an overcomplete dictionary with extreme sparsity, producing localized features with low entanglement. This sparsity makes oscillatory pair identification via the Hilbert-based procedure (Section~\ref{sec:pairs}) substantially cleaner: the selected pairs isolate the vortex-shedding mode without inadvertently coupling to boundary-layer dynamics, pressure gradients, or other non-oscillatory physics.

PCA directions are decorrelated but dense---every component is a global linear combination of all embedding dimensions---so rotating a PCA pair inevitably perturbs unrelated flow information. Raw MGN embeddings are fully entangled along both axes (neither sparse nor decorrelated), leaving essentially no room for targeted intervention.

The rotation mechanism completes the picture: it couples the right representation with the right intervention design, preserving the amplitude--phase structure of oscillatory modes while applying smooth, time-varying corrections. Neither ingredient alone is sufficient: SAE space without rotation (i.e., static interventions) fails, and rotation without SAE (i.e., in PCA or raw space) underperforms. The combination is what makes phase steering effective.
\section{Limitations and Future Work}
\label{sec:discussion}

This study focuses on a controlled proof of concept: the experiments are restricted to a single cylinder-wake regime and one target phase shift, so broader claims about steering scientific surrogates require evaluation across additional flow regimes, geometries, and surrogate architectures. In addition, the current Hilbert-based quadrature filtering is best suited to dynamics dominated by a single periodic mode and may be less reliable for multi-frequency or chaotic flows. Finally, steering operates within the manifold learned by the base MGN and therefore cannot recover physics missing from the underlying surrogate. Natural next steps include extending the framework to turbulent, multi-frequency regimes where multiple oscillatory modes must be steered simultaneously, validating across diverse geometries and surrogate architectures, and deploying the pipeline in online synchronization settings where steering parameters are updated from sparse sensor streams in real time. Investigating alternative pair identification strategies that extend beyond single frequency Hilbert analysis, such as wavelet methods or data driven mode decomposition, could further broaden applicability to flows with broadband or intermittent dynamics.


\section{Conclusion}
 
We have presented a post-hoc phase-steering framework that corrects temporal misalignment in frozen graph-based CFD surrogates by identifying near-quadrature oscillatory feature pairs and applying smooth, time-varying rotations in a low-rank coefficient space. The framework is representation-agnostic by design: the same pipeline was applied identically in SAE, PCA, and raw MGN embedding spaces, isolating representation quality as the key variable. On cylinder wake flow, SAE-based steering substantially outperformed both alternatives, while standard static latent interventions (scaling, additive perturbation, and clamping) failed to provide useful correction, demonstrating that techniques effective for steering language and vision models do not transfer directly to time-dependent physical surrogates. These results establish that effective steering in this setting requires two ingredients simultaneously: a sparse, disentangled representation that isolates oscillatory structure from unrelated flow physics, and an intervention mechanism that preserves the temporal coherence of that structure. More broadly, this work suggests that adapting SAE-based interpretability tools to scientific domains requires coupling the learned representation with domain-specific intervention design, here grounded in classical signal analysis and modal decomposition from fluid mechanics. The framework is modular: advances in SAE architectures, surrogate models, or physics-informed optimization can each be incorporated independently, offering a pathway toward steerable, interpretable surrogates for deployment in digital twins and closed-loop flow control.

\begin{ack}
This work was performed under the auspices of the U.S. Department of Energy by Lawrence Livermore National Laboratory under Contract DE-AC52-07NA27344. 
The work is partially funded by LDRD: 23-ERD-029, as well as DOE ECRP 51917/SCW1885.
This work is reviewed and released under LLNL-JRNL-2015715.

\end{ack}

\bibliographystyle{named}
\bibliography{ijcai25}





\end{document}